# Sensitive Room-Temperature Terahertz Detection via Photothermoelectric Effect in Graphene

Xinghan Cai[1], Andrei B. Sushkov[1], Ryan J. Suess[2], Mohammad M. Jadidi[2], Gregory S. Jenkins[1], Luke O. Nyakiti[4], Rachael L. Myers-Ward[5], Shanshan Li[2], Jun Yan[1,6], D. Kurt Gaskill[5], Thomas E. Murphy[2], H. Dennis Drew[1], Michael S. Fuhrer[1,3]

[1]Center for Nanophysics and Advanced Materials, University of Maryland, College Park, MD 20742-4111 USA; [2]Institute for Research in Electronics and Applied Physics, University of Maryland, College Park, MD 20742 USA; [3]School of Physics, Monash University, 3800 Victoria, Australia; [4]Texas A&M University, Galveston, TX 77553; [5]U.S. Naval Research Laboratory, Washington, DC 20375, USA; [6]Department of Physics, University of Massachusetts, Amherst, MA 01003, USA

Terahertz (THz) radiation has uses from security to medicine[1]; however, sensitive room-temperature detection of THz is notoriously difficult[2]. The hot-electron photothermoelectric effect in graphene is a promising detection mechanism: photoexcited carriers rapidly thermalize due to strong electron-electron interactions[3,4], but lose energy to the lattice more slowly[3,5]. The electron temperature gradient drives electron diffusion, and asymmetry due to local gating[6,7] or dissimilar contact metals[8] produces a net current via the thermoelectric effect. Here we demonstrate a graphene thermoelectric THz photodetector with sensitivity exceeding 10 V/W (700 V/W) at room temperature and noise equivalent power less than 1100 pW/Hz$^{1/2}$ (20 pW/Hz$^{1/2}$), referenced to the incident (absorbed) power. This implies a performance which is competitive with the best room-temperature THz detectors[9] for an optimally coupled device, while time-resolved

**measurements indicate that our graphene detector is eight to nine orders of magnitude faster than those[7,10]. A simple model of the response, including contact asymmetries (resistance, work function and Fermi-energy pinning) reproduces the qualitative features of the data, and indicates that orders-of-magnitude sensitivity improvements are possible.**

Graphene has unique advantages for hot-electron photothermoelectric detection. Gapless graphene has strong interband absorption at all frequencies. The electronic heat capacity of single-layer graphene is much lower than in bulk materials, resulting in a larger change in temperature for the same absorbed energy. The photothermoelectric effect has a picosecond response time, set by the electron-phonon relaxation rate. [10,11]. Hot electron effects have been exploited in graphene for sensitive bolometry in THz and millimeter-wave at cryogenic temperatures, by using temperature-dependent resistance in gapped bilayer graphene[12], which is sizable only at low temperature, or noise thermometry[13], which requires complex RF electronics. In contrast, our photothermoelectric approach is temperature insensitive and produces an observable dc signal even under room temperature conditions.

To realize our graphene hot electron thermoelectric photodetector we generate an asymmetry by contacting graphene with dissimilar metals using a standard double-angle evaporation technique as shown in Figs. 1a-e (also see Methods). Fig. 1f shows optical and atomic-force micrographs of our monolayer graphene device. Two metal electrodes, each consisting of partially overlapping Cr and Au regions, contact the monolayer graphene flake. The 3 μm × 3 μm graphene channel is selected to be shorter than the estimated electron diffusion length[14]. Fig. 1g shows the schematic of our detector in cross section. Figs. 1h-k illustrate the principle of operation: Electrons in graphene are heated by the incident light and the contacts serve as a heat sink, resulting in a non-uniform electron temperature $T(x)$ as a function of position $x$ within the device (Fig. 1h). Due to different metal contacts, the Fermi energy profile (Fig. 1i) and thus the Seebeck coefficient (*S*; Fig 1j) are asymmetric across the device. Diffusion of hot electrons creates a potential gradient $\nabla V(x) = -S\nabla T(x)$ (Fig. 1k). The total signal is the integral of $\nabla V(x)$ over the device length (area under the curve in Fig. 1k), and is non-zero because of the asymmetry.

Fig. 2 shows the responsivity *R*, the ratio of signal voltage to the absorbed power, of the device to dc or ac Joule heating, near infrared (IR; 1.54 µm), and THz (119 µm) excitation (see Methods). In order to better compare the response across such disparate wavelengths we define the responsivity using the absorbed power, rather than incident power. Our device absorbs only a small fraction of the incident THz power (estimated from the measured sheet conductivity; see Methods and Supplementary Note 3), however the absorption could in principle be increased by using multilayer graphene, using an antenna, or tailoring a plasmonic resonance in graphene to match the incident frequency. Thus results referenced to absorbed power highlight the ultimate potential for our device scheme. However, as we discuss below, even our unoptimized device with no antenna has performance referenced to *incident* power that is unrivaled in its combination of speed and sensitivity. Fig. 2a shows the two-probe conductance *G* as a function of gate voltage $V_g$ measured from the point of minimum conductance $V_{g,min}$. The effective charge carrier mobility is 1,500 cm²/Vs, likely an underestimate of the true mobility due to inevitable contact resistance in the two-probe geometry. Fig. 2b and 2c plot the responsivity $R(V_g)$ as a function of gate voltage for dc Joule heating and THz excitation, respectively. For both excitations, the peak responsivity appears at low carrier density, changes sign at $V_g - V_{g,min}$ = -20V and is small at large negative $V_g$. The overall shape and magnitude are comparable, suggesting that both signals are generated from the same mechanism – the hot carrier thermoelectric effect. The THz responsivity is slightly larger than dc, possibly reflecting a slight overestimation of the THz absorption due to (1) neglected contact resistance in estimating graphene's conductivity or (2) inhomogeneity, which causes the average conductivity to be greater than the inverse of the average resistivity. At a later time (150 days) we measured the conductance and responsivity to ac Joule heating and near IR illumination of the same device, shown in Figs. 2d-f. The device has degraded slightly showing somewhat higher $V_{g,min}$ and a slightly lower conductance. The responsivity under Joule heating (Fig. 2e) is also lower than previously measured (Fig. 2b) but shows similar functional form. The near IR responsivity is much lower than the far IR responsivity, possibly indicating the importance of optical phonon emission[5] in hot carrier relaxation for excitation energies exceeding the optical phonon energy (~160 meV). The near IR responsivity shows a different gate-voltage dependence, possibly due to contribution of the photovoltaic effect[6,15]. However, Fig. 2b, c, e, f together show that the thermoelectric signal persists from dc to near infrared frequency with comparable responsivity, implying that the photothermoelectric effect is a promising mechanism for extraordinarily broadband detection of radiation.

Fig. 3a shows the gate-voltage-dependent responsivity for a similar device; the peak responsivity to THz excitation is 715 V/W. Fig. 3b shows the measured gate voltage-dependent noise with no THz excitation (black dotted line) and the calculated Johnson-Nyquist noise floor $(4k_BT/G)^{1/2}$ (red dotted line), where $k_B$ is the Boltzmann constant and $G$ is the measured conductance. The experimental noise only slightly exceeds the theoretical limit, indicating that nearly Johnson-Nyquist noise-limited performance is attainable. As shown in Fig. 3c, the noise equivalent power (NEP) reaches a minimum level of 16 pW/(Hz)$^{1/2}$ at peak responsivity.

We now characterize the response time of our detectors. We first investigate the intrinsic time response of the devices using a pulse-coincidence technique[7,10] with a 1.56 μm pulsed laser (see Methods). Fig. 4a shows the photovoltage signal measured on another device similar to the one shown in Fig. 1 due to pump and probe beam as a function of the probe delay time $\tau_d$ at the temperature $T$ = 150 K. The dip of the signal at zero delay comes from nonlinearity in photoresponse at low temperature[7,10]. By fitting the data to a two-sided exponential decay (red line in Fig. 4a) we estimate an intrinsic response time of 10.5 ps, due to electron-phonon relaxation. We also fabricated detectors using dissimilar metal electrodes to contact epitaxial single-layer graphene on (0001) semi-insulating SiC (see Fig. 4d, and Methods) and large-area chemical vapor deposition-grown (CVD) graphene on $SiO_2$/Si (see Fig. 4e, and Methods), realizing devices capable of direct readout at microwave frequencies. Fig. 4b shows the time-domain response of the epitaxial graphene device to ultrafast optical (800 nm wavelength) pulses at room temperature and Fig. 4c shows the response of the CVD graphene device to ultrafast THz (0-2 THz) pulses, recorded by a 40 GHz oscilloscope (see Methods). The CVD graphene device active area is 500 μm square (Fig. 4e) to collect more incident power, and the $SiO_2$/Si substrate enabled a gate-dependent photoresponse measurement. Fig. 4c shows the differential response at $V_g$ = -40 V subtracted from the response at $V_g$ = -20V and $V_g$ = 0V to eliminate any gate voltage-independent background. As shown in Fig. 4b, FWHM (full width at half maximum) of the signal is 30 ps for 800 nm optical excitation. As the response is convolved with the 25 ps response of the oscilloscope itself, we conclude that the response time is significantly less than 30 ps, and consistent with the intrinsic 10.5 ps response time estimated in Fig. 4a. The electrical impulse response to THz excitation is 110 ps (Fig. 4c) which is slower because of the larger size (and capacitance) of the CVD device. Our results are consistent with other direct measurements of graphene's thermal response time in the near IR[11,16] and THz[17] where the characteristic time scale was found to be 10 - 100 ps.

We now compare our device to existing technologies. The NEP of our device, 16 pW/(Hz)$^{1/2}$ referenced to absorbed power is competitive with the best room-temperature low-frequency THz detectors[9]. However a significant advantage of our device is its speed. Graphene based room-temperature terahertz detectors based on a transistor geometry[17-20] have shown sensitive detection at 358 GHz[19], however our device's responsivity and NEP referenced to *incident* power are still superior to these devices. We anticipate room for two orders of magnitude sensitivity improvement by increasing absorption through e.g. antenna coupling, and further orders-of-magnitude improvements from increasing the thermopower asymmetry as discussed below. For frequencies above 1 THz, our reported responsivity is 5-6 orders of magnitude larger than in earlier graphene-based detectors [17,20], in part because photothermoelectric detection does not suffer from the high-frequency roll-off that is characteristic of FET-based detectors. Beyond graphene, there are few existing THz detector technologies with sub-100 ps response times. Schottky diodes can detect 100 ps signal modulations[21], but their responsivity decreases rapidly ($1/f^2$) with frequency $f$, and measured NEP are 0.3-10 nW/Hz$^{1/2}$ at 1 THz, increasing rapidly above 1 THz. An intraminiband superlattice detector[22] achieved a response time of 20 ps but responsivity was 50 µA/W (2.5 mV/W assuming 50 Ω load) at 6 THz, and a nanosize field-effect transistor[23] demonstrated 30 ps response at 5 THz with an estimated NEP >10 µW/Hz$^{1/2}$. Thus we believe our detector uniquely offers fast, sensitive detection in the few-THz regime, with orders of magnitude improvement in responsivity and NEP compared to existing THz detectors with sub-100 ps response times.

We now estimate the magnitude of the thermoelectric responsivity $R$, theoretically. First we ignore the electron-acoustic phonon coupling[14,24] and make a simple estimate based on diffusive cooling by the electrodes. According to the Wiedemann-Franz law and Mott relation[25, 26], graphene's electron thermal conductivity is $\kappa = L\sigma T$ and Seebeck coefficient is $S = LT(d\ln\sigma/dE_F)$, where $\sigma$ is the conductivity and the Lorentz number $L = \pi^2 k_B^2/3e^2$. A thermal difference $\Delta T$ results in a voltage $V = -S\,\Delta T$ and heat flux $Q = \kappa\,\Delta T$. Then $R = |V/Q| = (1/\sigma E_F)(d\ln\sigma/d\ln E_F) \approx 2/\sigma E_F$. The responsivity is maximized at small $E_F$ and small $\sigma$. These quantities are limited by disorder; for graphene on SiO$_2$ the minimal values are roughly $\sigma$ = 0.2 mS and $E_F$ = 50 meV[27], giving a maximum responsivity of $2\times10^5$ V/W, which is three orders of magnitude larger than our experimental result.

Next we model the response of our device considering three sources of asymmetry and qualitatively obtain their influence on the thermoelectric signal. We consider two effects in the models:

(1) asymmetry due to the contact metals, including pinning of the chemical potential at the graphene/metal interface and the long-ranged electrostatic effect of the nearby metal on graphene due to their different work functions[28], and (2) asymmetry in contact resistance[29]. The first effect is inevitable in our dissimilar-metal contacted devices. Additional scattering in graphene caused by metal near the contact may contribute to additional contact resistance[29] and it is reasonable to suppose that this effect may be asymmetric for different contact metals. See Supplementary Note 4 for details of the models.

Fig. 5 summarizes the results of the modeling, where we have used realistic parameters for gold and chromium metals in modeling the contact chemical potential pinning and workfunction[28], and an additional contact resistance of $R_c$ = 33.5 Ω for the gold electrode. In general we find that asymmetry in contact metal produces a signal symmetric in $|V_g - V_{g,min}|$ (Fig. 5a) while additional contact resistance produces a signal antisymmetric in $|V_g - V_{g,min}|$ (Fig. 5b). The combined effect of contact metal and contact resistance asymmetry (Fig. 5c) describes well the magnitude and the shape of the gate-voltage dependent response to THz excitation in the real device (replotted in Fig. 5d). We can identify the overall asymmetry as arising from contact resistance, and the dip in responsivity near charge neutrality as due to contact work function/Fermi-energy pinning effects. The model has several adjustable parameters (see Methods), and verification will require more work to systematically vary these and observe their effect on responsivity. However the fact that we can model the data with physically reasonable parameters indicates that model captures the essential operating principles of the device. We note that the responsivity is several orders of magnitude smaller than the maximum thermopower that might be expected for local heating of a *pn* junction. This suggests significant improvements of room-temperature graphene THz detectors are possible using local gates or locally-doped regions to define *pn* junctions.

METHODS

Single layer graphene is exfoliated from bulk graphite onto a substrate of 300 nm $SiO_2$ over low doped Si (100 - 250 Ω • cm). See Supplementary Fig. 1 for a Raman spectrum of the graphene used in the device in Fig. 1f. Dissimilar metal contacts are fabricated in one lithographic step using a tilted-angle shadow evaporation technique. The evaporation mask is fabricated using a standard electron-beam lithography technique using a bilayer resist [methyl methacrylate (8.5%)/methacrylic acid copolymer (MMA), Micro Chem Corp.; and poly(methy methacrylate) (PMMA), Micro Chem Corp.][30]. 20 nm chromium and 20 nm gold are deposited at different evaporating angles.

The dc thermoelectric responsivity is characterized by applying a dc voltage across the electrodes and measuring the resulting current $I_1 = I + I_{thermal}$ and $I_2 = -I + I_{thermal}$ under both polarities of the applied voltage $\pm V$, where $I$ is the current generated by the bias voltage and $I_{thermal}$ is the thermoelectric current. The applied voltage is 0.2 V and the Joule heating power is tens of microwatts. The thermoelectric responsivity is then $R = V_{thermal}/P = I_{thermal}/I^2 = 2(I_1+I_2)/(I_1-I_2)^2$. We verify that $I_{thermal}$ is much less than $I$ in the measurement. Similarly for low-frequency ac excitation, a bias current $I_{ac}(t) = I_0\sin(\omega t)$ at frequency $\omega$ = 15.7 Hz is applied to the device. Measurements are made in the regime where the thermoelectric voltage is much smaller than $V_0$, the amplitude of the applied voltage. The observed thermoelectric voltage $V(t)$ is proportional to the absorbed power, $P(t) = (GV_0^2/2)[1 - \cos(2\omega t)]$ where $G$ is the conductance. This second harmonic component of the voltage $V_{2\omega}\cos(2\omega t)$ is detected by a lock-in amplifier giving the responsivity $R = 2GV_{2\omega}/(I_0^2)$. For optical excitation, we uniformly illuminate the device with chopped continuous wave laser and detect the open-circuit photovoltage signal by using a voltage preamplifier and lock-in amplifier. The wavelength is 1.54 μm for the near infrared laser and 119 μm for the THz laser generated by $CO_2$-laser-pumped methanol gas. We measured five devices and all show a similar gate dependent photoreponse. To calculate the absorbed power under far infrared excitation, we performed scanning photovoltage measurement to characterize the beam profile and determine the incident power intensity on the graphene area (see Supplementary Fig. 3). We treat the device as a conducting layer sandwiched by air and silicon substrate to find the real electric field at the graphene layer, and consider Drude absorption to estimate the quantum efficiency (see Supplementary Note 3). All the measurements mentioned above are performed under ambient conditions at room temperature. For the noise measurement in Fig. 3b the gate voltage-dependent noise is measured with a lock-in amplifier at the frequency $f$ = 331 Hz, the same frequency as used to chop the THz laser for the responsivity measurement in Fig. 3a.

The intrinsic speed of our graphene photothermoelectric detectors was measured using the asynchronous optical sampling (ASOPS) method[31] with an ultrafast pulsed laser with wavelength 1.56 μm, pulse width ~60 fs and average power 50 mW as pump and probe sources with maximum scan length 10 ns and scan resolution ~100 fs. The sample was mounted in an optical cryostat at 150 K. The photovoltage was measured as a function of the pump-probe delay time. Additionally, we prepared devices suitable for direct time-domain measurement of their extrinsic response time using the same tilted-angle shadow evaporation technique. For optical (800 nm) excitation, the starting material was epitaxial single-layer graphene on (0001) semi-insulating (resisitivity > $10^9$ Ω-cm) SiC; see[32] for additional details. The semi-insulating SiC substrate eliminated stray capacitance of device to substrate and absorption of the incident light by the substrate. The graphene channel was 4 μm long and 100 μm in width as shown in Fig. 4d. The pads were contacted by a three-tip radio-frequency ground-signal-ground probe. The photoresponse was excited by a pulsed laser beam with wavelength 800 nm, pulse width ~50 fs, repetition rate 1 kHz and pulse energy of 250nJ. The device for THz excitation is fabricated using CVD grown single-layer graphene on a substrate of 300 nm $SiO_2$ over low doped Si (100 - 250 Ω • cm). As shown in Fig. 4e, many graphene channels were connected in series to enhance the signal. Each graphene channel was 4 μm long and 500 μm in width. Broadband terahertz pulses with a duration

~1 ps and a spectrum spanning 0-2 THz were produced through optical rectification of femtosecond pulses in a lithium niobate prism[33], and focused onto the device through a polymethylpentene (TPX) lens. The focused THz pulses had a beam diameter of approximately 1 mm and a pulse energy of 160 nJ at a repetition rate of 1 kHz. The output signal was recorded using a high speed (bandwidth = 40 GHz) sampling oscilloscope.

AUTHOR CONTRIBUTIONS

X. Cai, A.B. Sushkov, J. Yan, T.E. Murphy, H.D. Drew, and M.S. Fuhrer conceived the experiments. X. Cai fabricated the graphene photodetectors. X. Cai, A.B. Sushkov, and G.S. Jenkins carried out the THz measurements. X. Cai, R.J. Suess, M.M. Jadidi and S. Li carried out the near-IR and pulsed laser measurements. X. Cai and J. Yan carried out the dc and ac transport measurements. L.O. Nyakiti, R.L. Myers-Ward and D.K. Gaskill synthesized the graphene on SiC. All authors contributed to the manuscript.

ACKNOWLEDGMENTS

This work was sponsored by the U.S. Office of Naval Research (award numbers N000140911064, N000141310712, N000141310865), the National Science Foundation (ECCS 1309750), and IARPA. M.S.F. was supported in part by an ARC Laureate Fellowship. The authors acknowledge useful discussions with Dr. Virginia D. Wheeler and Dr. Chip Eddy, Jr.

ADDITIONAL INFORMATION

Supplementary information accompanies this paper at www.nature.com/naturenanotechnology. Reprints and permission information is available online at

http://npg.nature.com/reprintsandpermissions/. Correspondence and requests for materials should be addressed to M.S.F.

**Figure Captions**

FIG. 1. Graphene photothermoelectric detector device fabrication and principle of operation. (a-e) Lithographic sequence used to produce the graphene terahertz detector. (a) A bilayer resist (MMA/PMMA; see Methods) is spun onto graphene on $SiO_2$/Si. (b) Resist is patterned by electron beam and developed. Successive angled evaporations of chromium (red arrows) (c) and gold (yellow arrows) (d) followed by liftoff produces a single-layer graphene device with dissimilar metal contacts on the opposing sides as shown schematically in (e). (f) Optical micrograph showing electrical contacts and (inset) atomic force micrograph showing bimetallic contacts connected to an exfoliated graphene layer. (g-k) Schematic of the principle components during device operation. (g) Cross-sectional view of the device. (h-j) Profiles across the device of (h) electron temperature $T(x)$, (i) Fermi level $E_F(x)$, (j) Seebeck coefficient $S(x)$ and (k) potential gradient $\nabla V(x) = -S\nabla T(x)$. The photoresponse is the integral of $\nabla V(x)$ over the length of the device, or area under the curve in (k).

FIG. 2. Broadband thermoelectric responsivity of graphene photothermoelectric detector. (a,d) Electrical conductance, (b,e) responsivity to Joule heating, and (c,f) responsivity to radiation as a function of gate voltage for the device shown in Fig. 1f at room temperature and in ambient environment. Data in panels (d-f) were taken 150 days after data in panels (a-c). In (a-c) the minimum conductivity point is $V_{g,min}$ = 42 V, and in (d-e) $V_{g,min}$ = 80 V. Responsivity to Joule heating was measured at dc in (b) and at 15.7 Hz using the second harmonic technique in (e) (see Methods). Panel (c) shows responsivity to 119 μm wavelength THz radiation referenced to the absorbed power and panel (f) shows response to 1.54 μm infrared radiation.

FIG. 3. Noise equivalent power of graphene photothermoelectric detector. (a) Responsivity to 119 μm wavelength THz radiation, (b) measured noise (black dotted line) and calculated Johnson-Nyquist noise (red dotted line), and (c) measured noise equivalent power (NEP) as a function of gate voltage for a similar device to the one shown in Fig. 1f. The blue line corresponds to NEP = 16 pW/Hz$^{1/2}$. The responsivity and NEP are referenced to the absorbed power. For clarity, NEP is plotted in log scale.

FIG. 4. Response time of graphene photothermoelectric detector. (a) Photoresponse from pump-probe laser pulses as a function of delay time at 150 K. Red solid line shows a best fit assuming exponential decay of hot-electron temperature. (b) Time domain photoresponse to pulsed laser excitation at 800 nm wavelength recorded by a 40 GHz sampling oscilloscope for device fabricated on SiC (see Methods). The FWHM response is ~30 ps. (c) Time domain photoresponse to pulsed laser excitation in THz range recorded by a 40 GHz sampling oscilloscope for a CVD graphene device (see Methods). The FWHM response is ~110 ps. Black (Red) line shows the response at $V_g$ = -20 V (0 V). The micrographs of the devices for the measurements in (b) and (c) are shown in (d) and (e) respectively.

FIG. 5. Simulated responsivity of graphene photothermoelectric detector. The assumed asymmetry of the device is (a) induced by the work function difference of Cr and Au and different chemical potential pinning near both contacts, (b) purely induced by an additional contact resistance near the Au electrode, (c) induced by the asymmetries shown in (a) and (b) together. (d) Measured responsivity of our device to 119 μm wavelength THz radiation (replotted from Fig. 2c).

Figure 1.

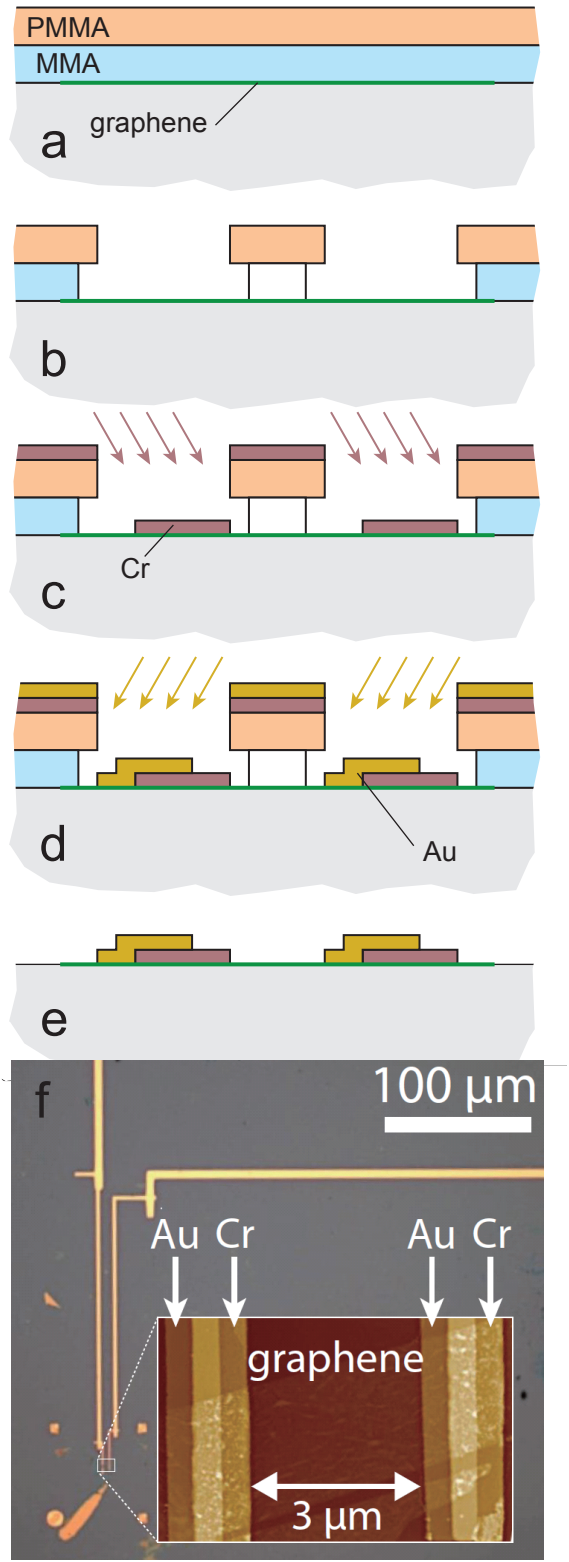
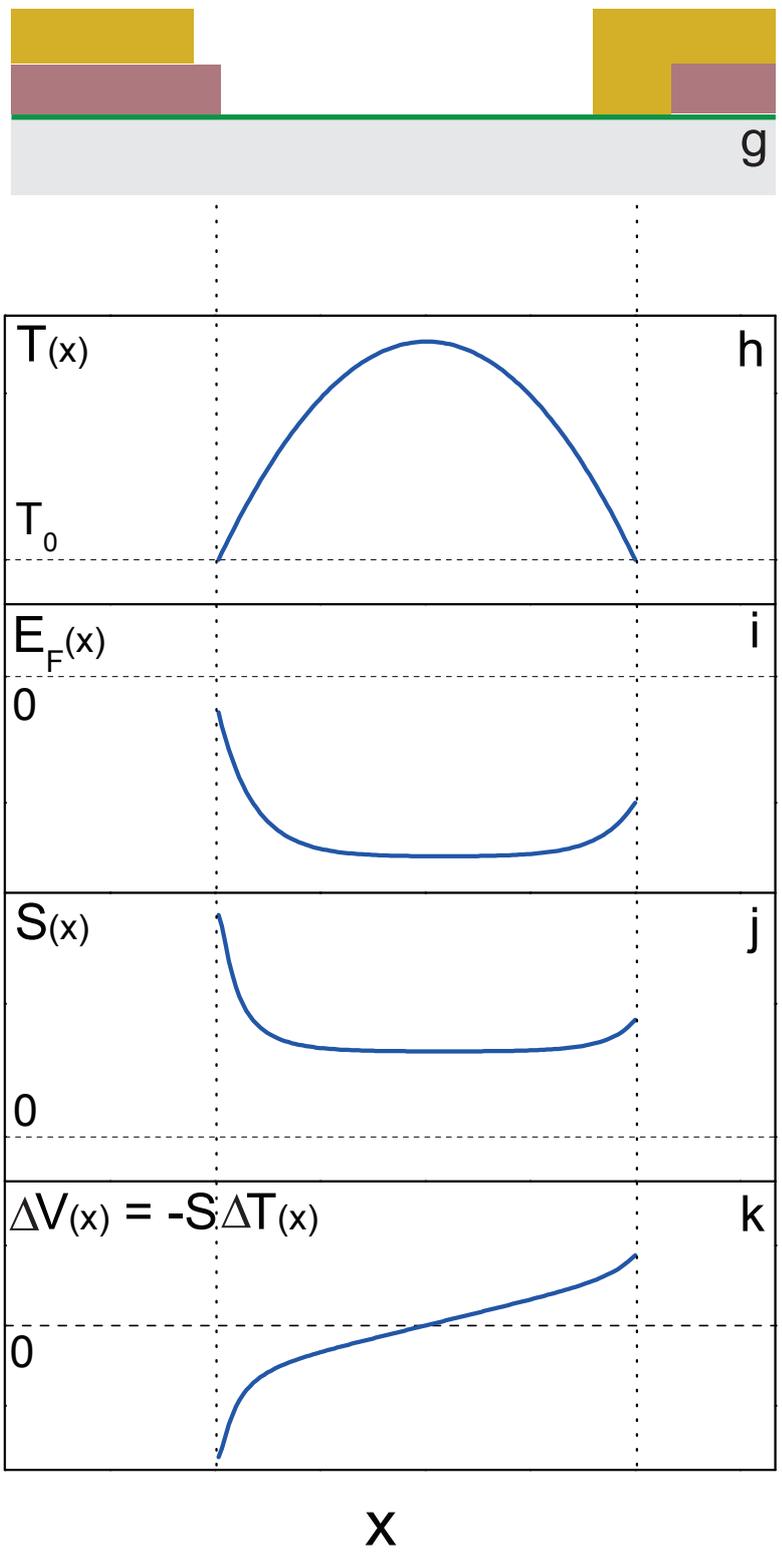

Figure 2.

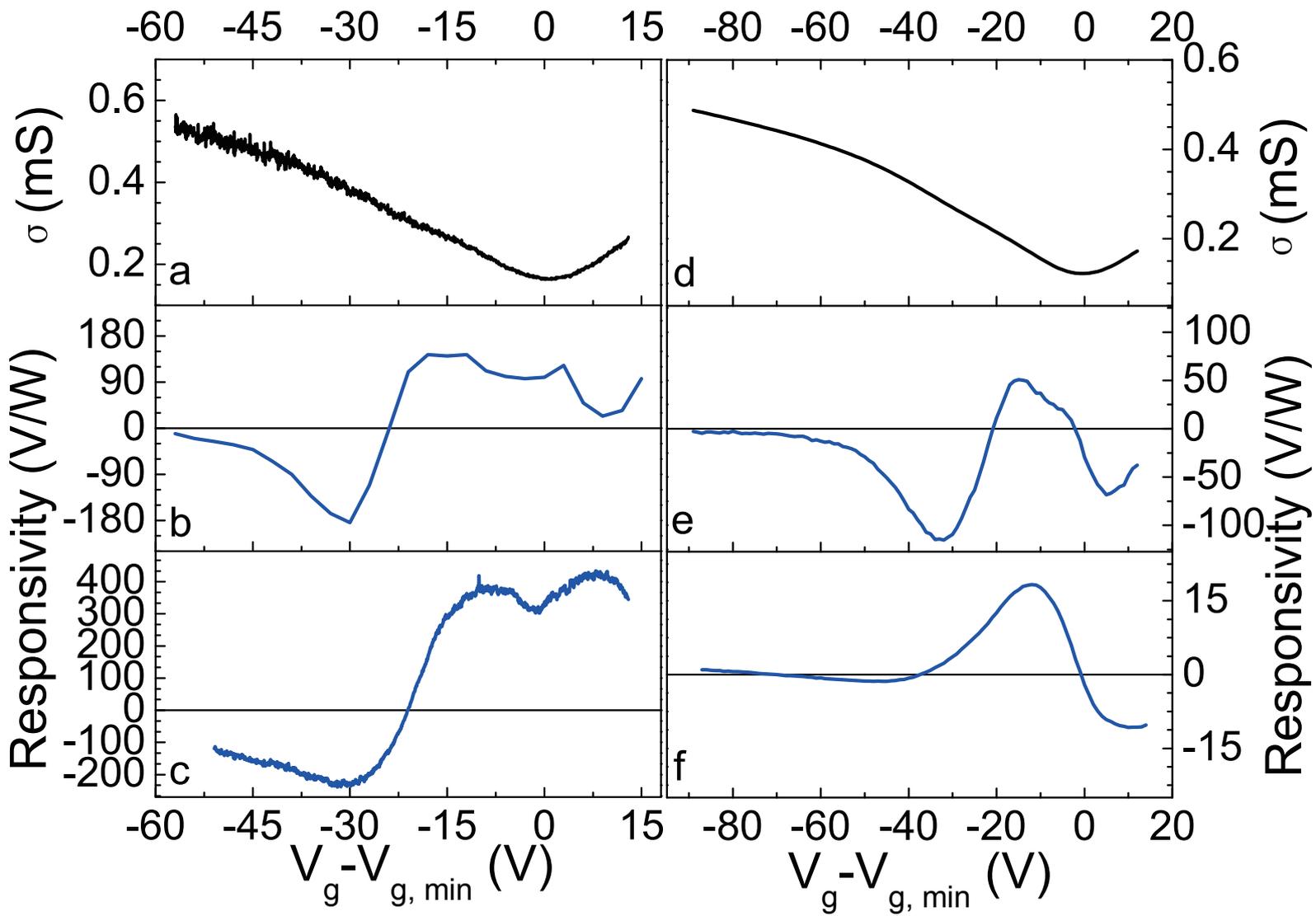

Figure 3.

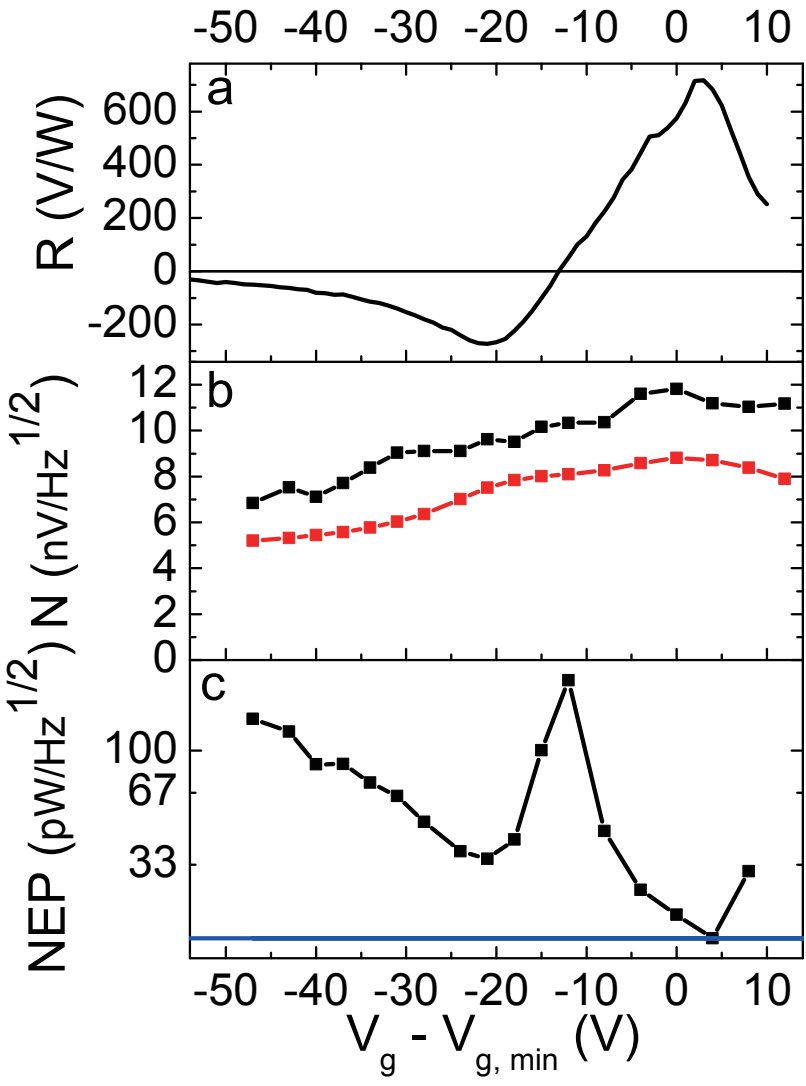

Figure 4.

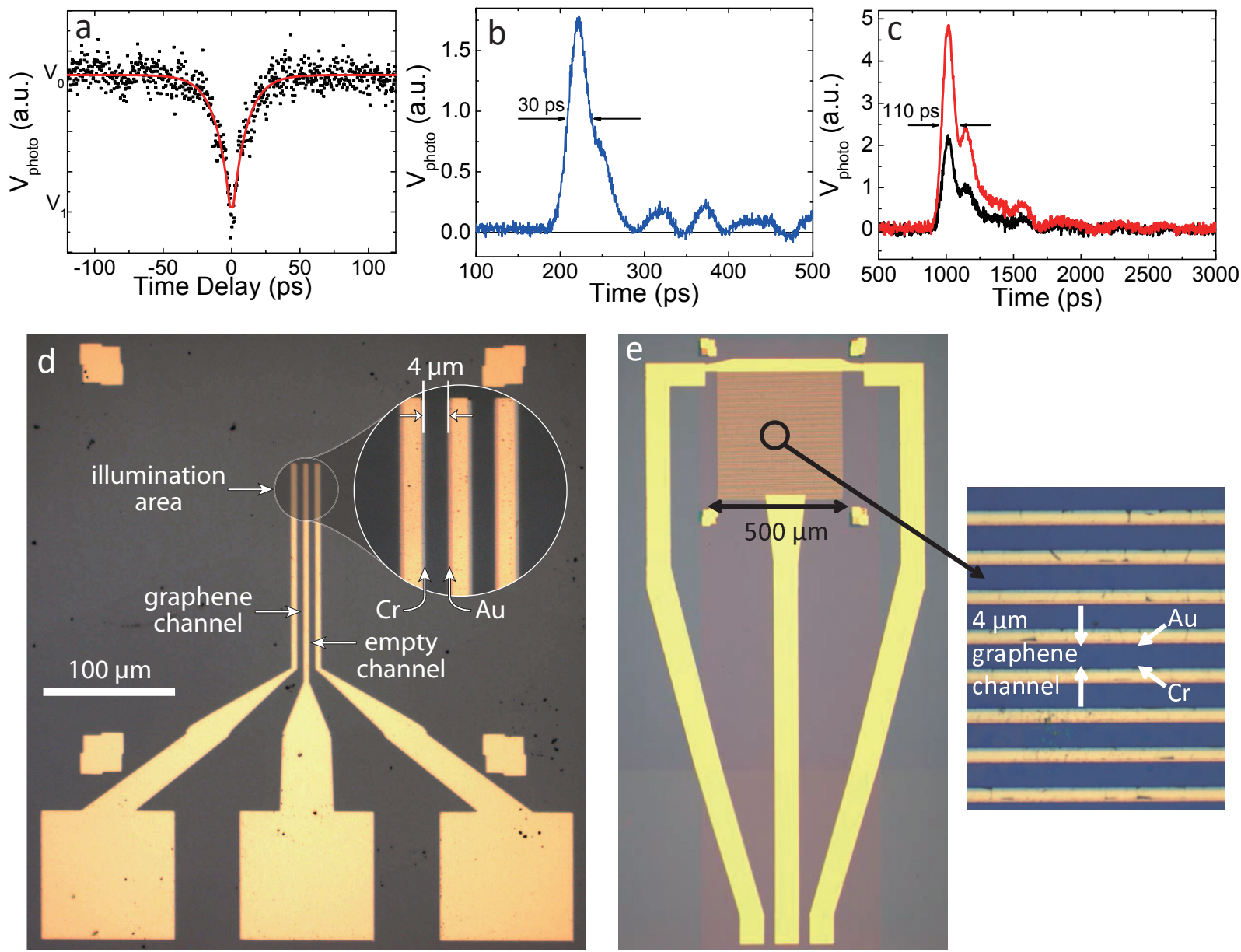

Figure 5.

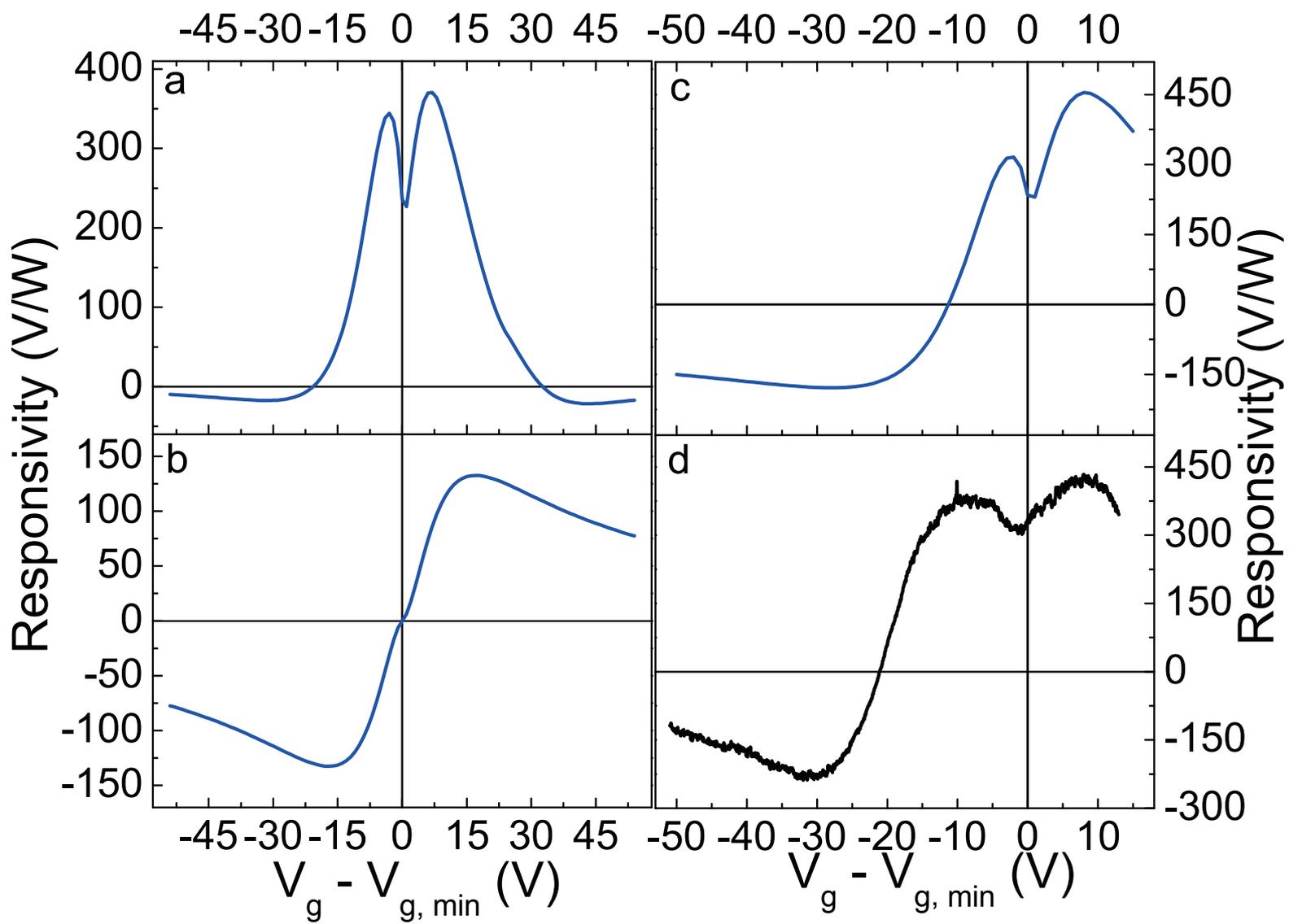

# Supplementary Information for manuscript

# "Sensitive Room-Temperature Terahertz Detection via Photothermoelectric Effect in Graphene"

Xinghan Cai[1], Andrei B. Sushkov[1], Ryan J. Suess[2], Mohammad M. Jadidi[2], Gregory S. Jenkins[1], Luke O. Nyakiti[4], Rachael L. Myers-Ward[5], Shanshan Li[2], Jun Yan[1,6], D. Kurt Gaskill[5], Thomas E. Murphy[2], H. Dennis Drew[1], Michael S. Fuhrer[1,3]

[1]*Center for Nanophysics and Advanced Materials, University of Maryland, College Park, MD 20742-4111 USA;* [2]*Institute for Research in Electronics and Applied Physics, University of Maryland, College Park, MD 20742 USA;* [3]*School of Physics, Monash University, 3800 Victoria, Australia;* [4]*Texas A&M University, Galveston, TX 77553;* [5]*U.S. Naval Research Laboratory, Washington, DC 20375, USA;* [6]*Department of Physics, University of Massachusetts, Amherst, MA 01003, USA*

## Supplementary Notes

**Supplementary Note 1: Raman spectroscopy of graphene**

We performed Raman spectroscopy on our exfoliated graphene used for the device shown in Fig. 1f and SiC graphene used for the device shown in Fig. 4b inset of the main text. The spectra are shown in Supplementary Fig. 1 (a)- (c).The single-Lorentzian 2D peak indicates single layer graphene in both cases and the near absence of the D peak in Supplementary Fig. 1 (a) shows that our exfoliated graphene's quality is high.

**Supplementary Note 2: Power dependence**

Since both the electron thermal conductivity and the Seebeck coefficient of graphene are proportional to the temperature according to the Wiedemann-Franz law and Mott relation, the thermal voltage, either generated by Joule heating or photon excitation, should be linearly dependent on the absorbed power. Supplementary Fig. 2 shows the power dependence of the response for Joule heating, near-IR and far-IR radiation at a fixed gate voltage. The data is taken on one device for Supplementary Fig. 2(a-b) and on another similar device for Supplementary Fig. 2(c). Red lines are linear fitting to the experimental result. Supplementary Fig. 2 shows that the voltage response is proportional to the absorbed power (i.e. the responsivity is independent of power) over a power variation of 3 orders of magnitude. We also measured gate-voltage dependent responsivity of the device at various applied powers and find the linear response happens at all gate voltages, verifying that the device is operating in the linear regime at room temperature, and that our assumption that the signal is generated by heating is correct.

**Supplementary Note 3: Absorbed power calculation**

In order to quantitatively analyze the responsivity of the device in the main text, we consider the responsivity to absorbed power instead of the total incident power. Here we show how we calculate the absorbed power of the device which shows a peak responsivity of 715 V/W. We first measured the power intensity distribution of the laser beam. Since our device's active area is much smaller than the laser's spot size, we can approximate the device as a point. As shown in Supplementary Fig. 3, when scanning the beam across the device, the spatial distribution of the photovoltage signal reflects the beam intensity profile. We fit the data using a Gaussian function:

$$V_{\text{photo}} = V_{\text{bg}} + \frac{V_0}{w\sqrt{\pi/2}} e^{-2r^2/w^2}$$

where $V_{bg}$ is the background signal due to electrical pick-up and other noise source and *r* is the distance to the center of the device. As shown in the inset of Supplementary Fig. 3, the graphene flake's size is ~ 2.0 μm x 2.1 μm. For convenience of calculation, we approximate its shape as a disk with the same area (radius $r_0 = 1.16$ μm). Considering the total incident power of 17 mW (The laser power was measured with a thermopile calibrated at NIST, Boulder.), the power on device's active area can be expressed as:

$$P_{active} = P_0 \cdot \frac{\int_0^{r_0} e^{-2r^2/w^2} r dr}{\int_0^{\infty} e^{-2r^2/w^2} r dr} = 0.75 \text{ μW}$$

In addition, monolayer graphene will absorb only a small fraction of the incident power. In the THz range, the absorption is mainly due to the Drude response and can be expressed as $P = \frac{1}{2} \text{Re}[\sigma(\omega)] |E_t|^2 A$, where $\sigma(\omega)$ is graphene's conductivity, $E_t$ is the electric field on the graphene of area A. The electric field on graphene is related with the electric field of the incident beam $E_0$ as $E_t = \frac{2E_0}{|1+n+Z_0\sigma(\omega)|}$, where n = 3.42 is the refractive index of silicon substrate and $Z_0 = 377\Omega$ is the impedance of free space. Using the equation of the incident light intensity $I_0 = \frac{E_0^2}{2Z_0}$, we can write the absorption rate of graphene as:

$$\eta = \frac{P}{P_0} = \frac{\text{Re}[\sigma(\omega)]|E_t|^2/2}{E_0^2/2Z_0} = \frac{4Z_0 \text{Re}[\sigma(\omega)]}{|1+n+Z_0\sigma(\omega)|^2}$$

For our wavelength $\lambda = 119$ μm, $\sigma(\omega)$ can be approximated as our measured dc conductivity $\sigma_0$. Taking into account that our maximum photovoltage signal is 8.1μV, the peak responsivity is then expressed as:

$$\text{Responsivity}^{max} = \frac{V_{photo}^{max}}{P_{active} \cdot \eta} = 715 \text{V/W}$$

**Supplementary Note 4: Device response modelling**

Modeling the device response was done as follows. The device was approximated as a 3 μm×3 μm square. We assume that the local electrical conductivity σ of graphene depends on the local Fermi energy $E_F$ as:

$$\sigma = \sigma_{\min}(1+\frac{E_F^{\,4}}{\Delta^4})^{1/2} \tag{S1}$$

where $\sigma_{\min}$ is the minimum conductivity and $\Delta$ is a parameter that expresses the disorder strength[S1]. This functional form for $\sigma$ correctly extrapolates between the highly doped region where $\sigma \sim E_F^2$ and the charge neutral point where $\sigma \sim$ constant. Supplementary Figure 4 reproduces the $G(Vg)$ data from Fig. 2a with a fit to Eqn. S1 (red curve) to obtain $\sigma_{\min}$ = 0.169 mS and $\Delta$ =107 meV. To treat asymmetry in contact metal we followed the results of reference [S2] to obtain the charge carrier distribution across the device and thus the local Fermi level. Then we numerically solve the 1D diffusive heat conductance equation to get the temperature profile across the device[S1]. Given the temperature profile and local Fermi level we calculate the thermoelectric field $E = S\nabla T$, where $S$ is given by Eqn. (S1) and the Mott relation $S = LT(d\ln\sigma/dE_F)$, and integrate over the device to obtain the thermoelectric voltage. For chromium and gold we select parameters $V_{b1}$ = 65 meV and $V_{b2}$ = 265 meV for gold, $V_{b1}$ = -67 meV and $V_{b2}$ = 65 meV for chromium according to the model in Reference[S2].

We treat the thermoelectric signal due to asymmetric contact resistance as follows. We assume that the whole device is uniformly doped with Fermi energy determined by the gate voltage, and add an extra contact resistance $R_c$ = 33.5 Ω to the region from the gold contact extending 100 nm inside the graphene (the corresponding contact resistivity is $\rho_c$ = 1000 Ω). Then, the conductivity of this region can be rewritten as

$$\frac{1}{\sigma} = \frac{1}{\sigma_{\min}(1+\frac{E_F^{\,4}}{\Delta^4})^{1/2}} + \rho_c.$$

The electron thermal conductivity and Seebeck coefficient in this region change correspondingly.

To model the combined effects of contact metal and contact resistance, we first calculate the Fermi level distribution taking into account the contact metal asymmetry. The temperature profile is calculated from the thermal conductivity assuming an extra contact resistance $R_c$ = 33.5 Ω in the region from the gold contact extending 100 nm inside the graphene. The local Seebeck coefficient and the thermopower are then calculated as before.

**Supplementary Figures**

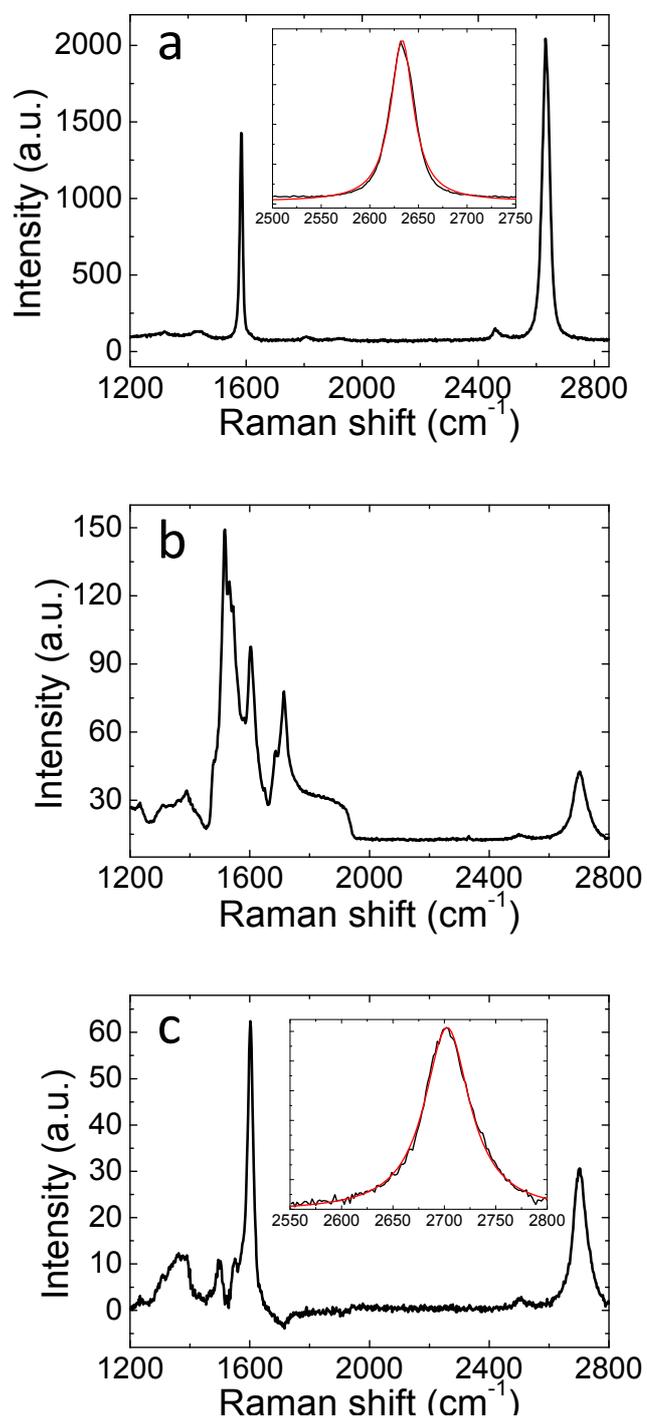

**Supplementary Fig. 1** Raman spectrum of graphene used for device shown (a) in Fig. 1f, main text (b) in Fig. 4b inset, main text (c) in Fig. 4b inset, main text (SiC background spectrum subtracted). The inset of (a) and (c) shows a Lorentzian fit (red line) to the 2D peak (black line) of corresponding spectrum.

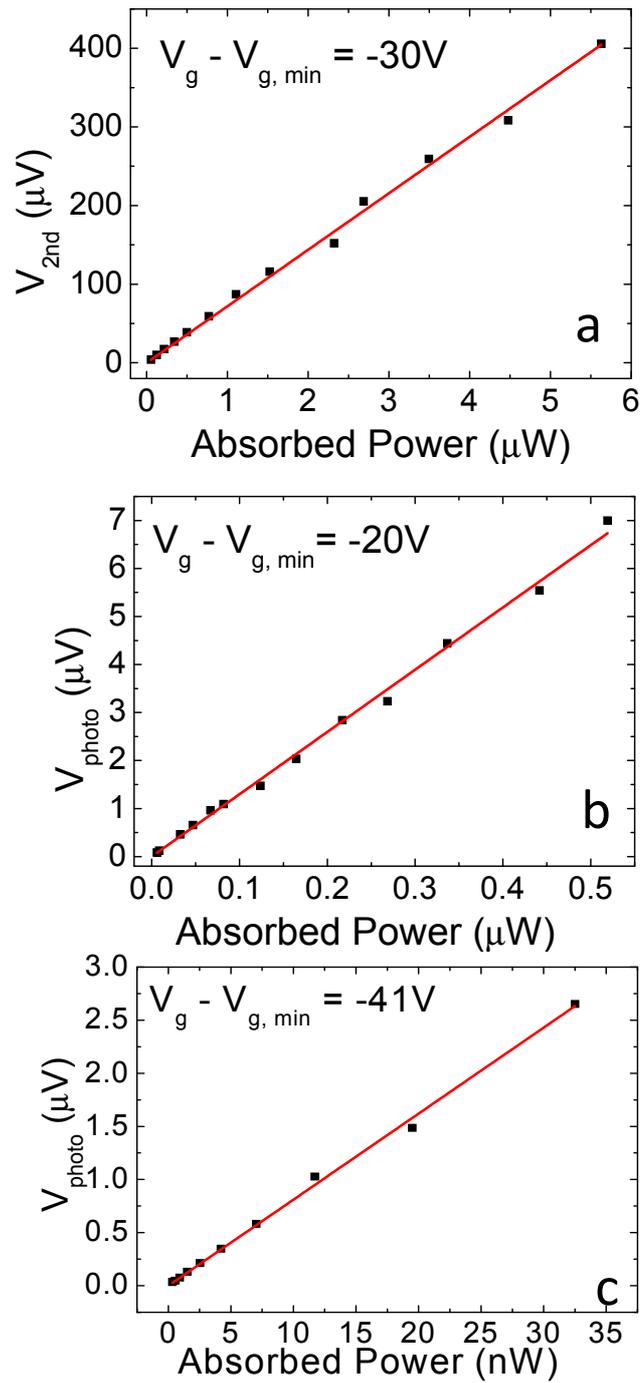

**Supplementary Fig. 2** Voltage signal as a function of applied power for (a) ac Joule heating at $V_g - V_{g,min}$ = -30 V, (b) 1.54 μm near infrared radiation at $V_g - V_{g,min}$ = -20V and (c) 119 μm far infrared radiation at $V_g - V_{g,min}$ = -41V.

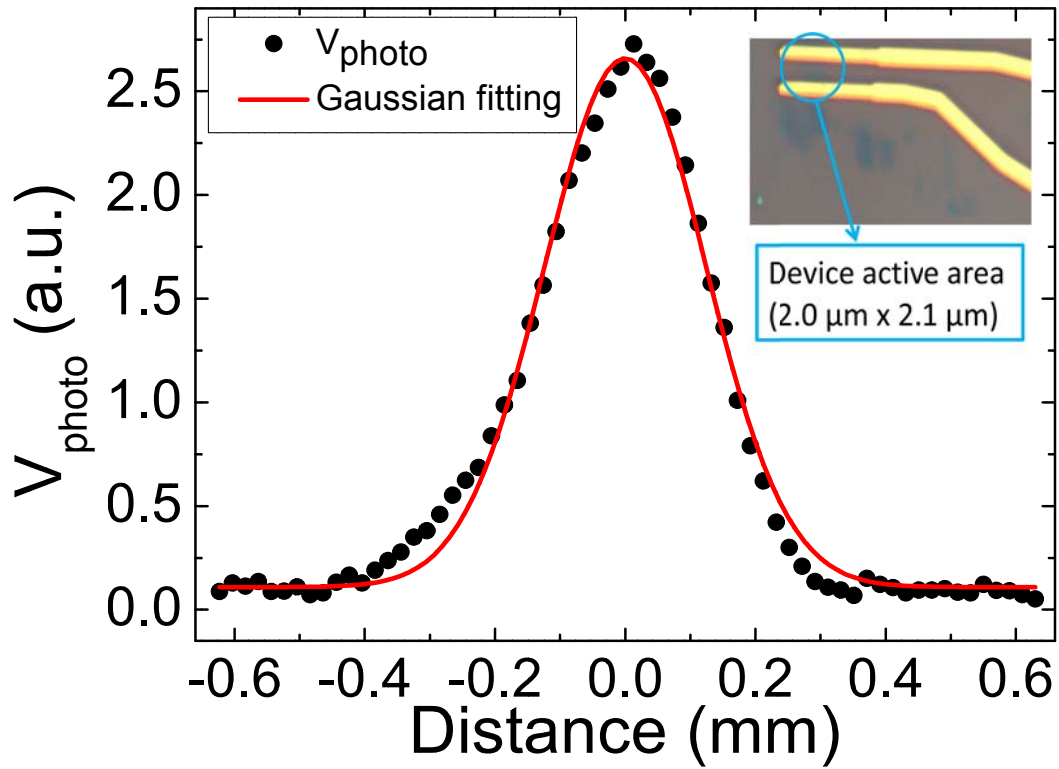

**Supplementary Fig. 3** Photovoltage for the graphene photothermoelectric detector as a function of distance measured as the far infrared laser is scanned across the device (black dots) and Gaussian fit to the experimental data (red curve). Inset: Optical micrograph of the device. The device active area (graphene flake) is between two metal electrodes.

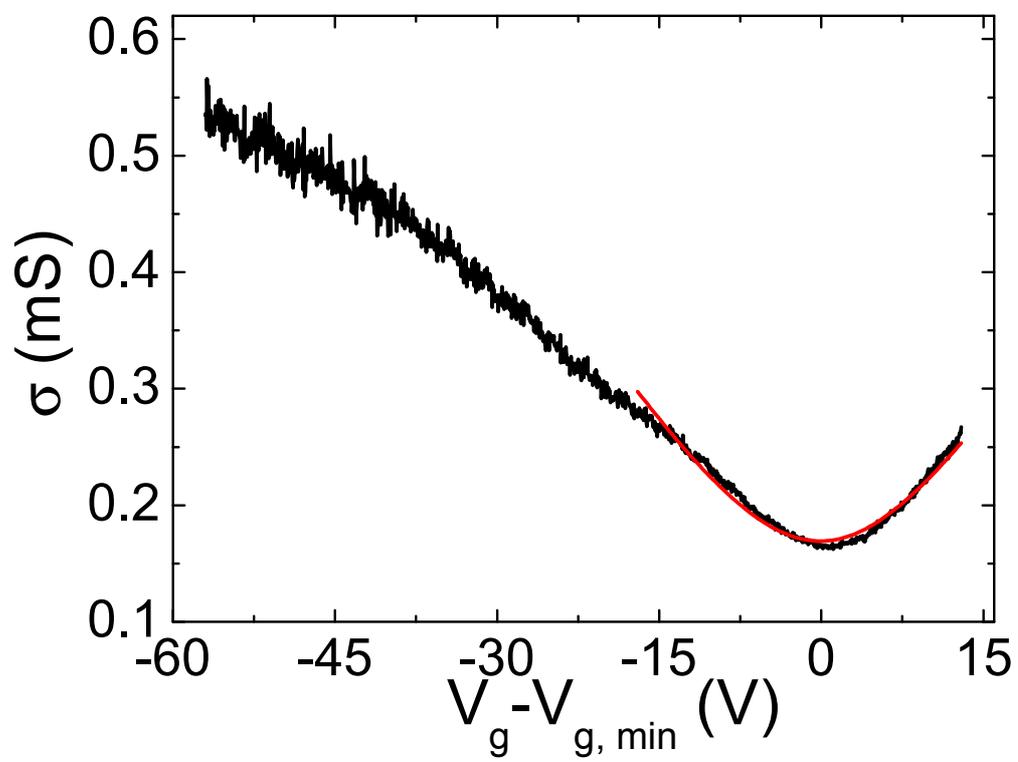

**Supplementary Fig. 4** Electrical conductance as a function of gate voltage (black curve) for the device shown in Fig. 1f. Red solid line is a fit to Eqn. S1.